\documentclass[a5paper, 11pt]{article}
\usepackage{graphicx}
\usepackage{fullpage}
\usepackage{fancyhdr}
\usepackage{enumerate}
\usepackage{mathtools}
\usepackage{amsmath}
\usepackage{txfonts}
\usepackage{gensymb}
\ProvideTextCommand{\DJ}{OT1}{\raisebox{0.25ex}{-}\kern-0.4em D}
\usepackage[titletoc]{appendix}

\usepackage[utf8]{inputenc}
\usepackage[swedish]{babel}
\usepackage{amssymb}
\usepackage[bottom=1.5cm,top=2.5cm,left=2.1cm,right=2.1cm]{geometry}
\usepackage{sectsty}
\usepackage[symbol]{footmisc}
\sectionfont{\large}

\addto\captionsswedish{}
\addto\captionsswedish{}
\addto\captionsswedish{}

\makeatletter
\def\@seccntformat#1{\csname the#1\endcsname.\quad}
\def\numberline#1{\hb@xt@\@tempdima{#1\if&#1&\else.\fi\hfil}}
\makeatother

\begin{document}
\begin{sloppypar}
\pagestyle{fancy}
\renewcommand{\headrulewidth}{0pt}
\fancyhf{}
\cfoot{\text{\thepage}}

\thispagestyle{empty}
\clearpage
\pagenumbering{arabic} 

\Large
\noindent
\begin{center}
\textbf {The Nature of the Chemonuclear Reaction}\footnote[2]{This article is an excerpt from Hidetsugu Ikegami’s book manuscript.} 
\end{center}
\vspace{1.5em}
\small
\noindent
\begin{center}
\text{Hidetsugu Ikegami}\\
\vspace{0.25em}
\footnotesize
\text{1930--2025}
\end{center}
\footnotesize
\noindent
\begin{center}
\text{Professor Emeritus, Osaka University}
\text{Formerly Director, Research Center for Nuclear Physics, Osaka University}
\text{Honorary Doctorate, Uppsala University}
\end{center}

\noindent
\begin{center}
\textit{Edited by}\\
\small
\text{Masako Ikegami}
\end{center}
\footnotesize
\noindent
\begin{center}
\text{Professor Emeritus, Institute of Science Tokyo/Tokyo Institute of Technology}
\text{Cooperative Researcher, Research Center for Nuclear Physics, Osaka University}
\text{E-mail: mikegami@rcnp.osaka-u.ac.jp}
\end{center}

\newpage
\large
\noindent
\setcounter{page}{1}
\begin{center}
\text{\textbf{1.} The Nature of the Chemonuclear Reaction}
\end{center}

\small
\noindent
\begin{center}
\text{Hidetsugu Ikegami and Masako Ikegami}
\end{center}

\Large
\hyphenchar\font=-1
\section*{}
\normalsize
\begin{abstract}
\noindent In the chemonuclear reaction, the bulk of itinerant s-electrons revealing thermodynamical liquid activity in metallic systems undergo contact interaction with atomic nuclei and nucleons, thereby inducing contagiously thermodynamical liquid activity among the bulk of nuclei towards the chemical potential minimum, hence the united spontaneous, irreversible atomic and nuclear reactions. The chemonuclear reaction is enhanced with astronomical figures beyond the enhancement of few particle processes (e.g., in the electron-screened pycnonuclear fusion). In the systems of metallike hydride/deuteride and electron donor mixtures, self-sustained H-H and D-D chemonuclear fusion reactions take place. Unforeseen phenomena are observed, e.g., radiationless fusion and $\gamma$-ray missing positron annihilation. In addition, the coherent $\mathrm{D_2}$-$\mathrm{D_2}$ and $\mathrm{D_3}$-$\mathrm{D_3}$ fusion reactions take place with enhancement factors over $10^{20}$$\sim10^{30}$ and $10^{30}$$\sim10^{46}$ respectively. Through electron density correction on the chemical potential of reactants, the enhancement of chemonuclear fusion approaches to that of enhanced pycnonuclear fusion in astrophysical condensed plasma in the white dwarf progenitors of supernovae. In essence, chemonuclear fusion in condensed matters revealing thermodynamical liquid activity is identical to enhanced pycnonuclear fusion in astrophysical condensed plasma. 

\vspace{.5em}
\noindent
\it{Keywords:} chemonuclear fusion; enhanced pycnonuclear fusion; metallic liquid; thermodynamical liquid activity
\end{abstract}

\section*{1.1 Introduction}
\normalsize
\noindent
Spontaneous chemical reactions (e.g., burning in the air or ionic reactions in liquids) are the most common irreversible phenomena. Certain ionic liquids may undergo rapid reactions with an enhancement factor near $10^{50}$ in the presence of thermodynamical liquid activity (i.e., Widom's general concept). In nuclear reactions in a metallic system with inbuilt thermodynamical liquid activity, the bulk of itinerant s-electrons undergo contact interaction with reactant nuclei, inducing contagiously thermodynamical liquid activity (i.e., macroscopical scale correlations) among the bulk of nuclei towards the chemical potential minimum of the system. With this “liquid activity contagion”, united atomic and nuclear reactions take place. We raise the issues of the coherent chemonuclear fusion, cascade enhanced nuclear reactions, the chemonuclear fission, and the stimulated emission of coherently line-up $\alpha$-cluster. These issues may lead to a major revision of the belief that nuclear reactions are independent of the atomic states of reactants. In a world of chemonuclear reaction, macroscopic scale nuclear reactions are subject to the chemical potential that governs the linked atomic or chemical reactions. Similar to super enhanced nuclear reaction in condensed plasma in stars (e.g., extraordinarily enhanced pycnonuclear fusion in supernova progenitors), the chemonuclear reaction will be the mostly enhanced nuclear reaction that we have seen on Earth.

For decades, astrophysicists predicted remarkable nuclear reaction rate enhancements in condensed plasma in stars through quantum statistical treatments. The enhancement with a factor of 30 orders magnitude was expected in metallic hydrogen liquid plasma in a white dwarf progenitor of a supernova. This kind of enhancement can be attributed to thermodynamical liquid activity revealed in dense itinerant s-electrons and reactant nuclei in metallic liquids. In 2001, Hidetsugu Ikegami found an astronomically enhanced $^7$Li(D, n)$^8$Be$\longrightarrow 2\cdot ^4$He reaction rate in a metallic Li liquid based on the microscopic consideration of the slow ion collision process. The rate enhancement $K(\mathrm{Be})=\mathrm{exp}[-\Delta G_\mathrm{r} \mathrm{(Be)}/k_\mathrm{B}\mathrm{T}]$ specified by chemical potential drop $-\Delta G_\mathrm{r}$(Be) in the intermediate united atom (quasi Be-atom) formation in the reaction was predicted. This predicted enormous rate enhancement was confirmed by detecting those $\alpha$-particles that have been produced during slow D-ions implantation on a Li-liquid target. This event signifies the existence of the chemonuclear reaction.

In this new scheme of reaction in a metallic Li liquid, even doubly intensified enhancement $K(\mathrm{Be}_2)=K^2(\mathrm{Be})=\mathrm{exp}[-2\Delta G_\mathrm{r} \mathrm{(Be)}/k_\mathrm{B}\mathrm{T}]$ was expected with the coherent formation of intermediate quasi-Be$_2$ molecule under the slow D$_2$-ion implantation. When investigating the function of Ni nanocrystal powder as a stabilizer for the Li liquid, Hidetsugu Ikegami found the unique characteristics of the system of hydrogen adsorbent Ni nanocrystal powder and Li and/or Li hydride mixtures. In the 1970s, metal scientists saw the possible formation of metallic hydrogen in adsorbents through systematic experiments with alloys. In their adsorbents, hydrogen or Li hydride molecules dissociate on the surface of adsorbent Pd or Ni nanocrystals. They are transmuted into metallic hydrogen and Li releasing itinerant s-electrons respectively. According to their quantum statistical treatments, the electrons are expected to induce a lower melting point of adsorbed metallic hydrogen. In the presence of dense itinerant s-electrons, this system reveals thermodynamical liquid activity among the bulk of reactant nuclei that may undergo astronomically enhanced nuclear reactions. As a result, the astronomically enhanced coherent D$_2$-D$_2$ and D$_3$-D$_3$ fusion reactions take place, thereby producing energetic (23.8 MeV) He ions. These He ions provoke diverse enhanced cascade chemonuclear reactions among the reactants charged in this system. This system has the features of the most active reactor that has ever been built, i.e., hybrid fusion/fission reactors, radioactive waste vanishing reactions, a variety of element synthesis systems. The chemonuclear reaction, taking place on a macroscopic scale, is subject to the chemical potential of the reaction.

Hidetsugu Ikegami spent two decades doing his chemonuclear reaction research. He surmised that enormously enhanced pycnonuclear fusion in condensed plasma in the white dwarf progenitors of supernovae might be achieved on Earth, especially in certain itinerant s-electron-rich liquid metallic media. The first evidence for his claim came from Uppsala University. The $^7$Li(D, n)$^8$Be$\longrightarrow 2\cdot^4$He chemonuclear reaction, together with the 2Li(D$_2$, 2n)$2\cdot^8$Be$\longrightarrow 4 \cdot ^4$He coherent chemonuclear reaction, was confirmed in Tokyo where the $^7$Li($^7$Li, 2n)$^{12}$C chemonuclear reaction was also observed. He further found that such hydrogen adsorbents as Ni and Pd nanocrystal powders are the most powerful stabilizer for metallic Li liquids that are extremely unstable even in a high vacuum. He began to wonder if the mixtures of the metallike hydride of these metals and such itinerant s-electron donors as Li and CaO form the most ideal systems of the chemonuclear reaction. This surmise was confirmed by experiment in Bologna through the efforts of Uppsala and Bologna research groups.

\hyphenchar\font=-1
\section*{1.2 Chemonuclear Reactions in Metallic Liquids}
\noindent
Nuclear fusion can be enormously enhanced in high-density matters in stars [1], hence the pycnonuclear (dense nuclei) reaction [2]. In condensed matters, electrons act to screen the Coulomb repulsion between the atomic nuclei. This screening effect is so remarkable that the rates of reactions at low temperatures are almost independent of the temperature and mostly dependent on the density of matters. The very cohesive effect through the solidification of dense liquids will enhance greatly the reaction rate [3, 4]. In condensed plasma in a white dwarf progenitor of a supernova, the rate of the nuclear reaction can be enhanced with a factor of an astronomical figure [5]. It should be noted that this rate enhancement in metallic liquids is mostly caused by the very cohesive effect. Although this enhancement is infeasible in gas plasma like the solar interior, it is common among spontaneous reactions in liquids.

By detecting those $\alpha$-particles that have been produced under some kV deuterium ion implantation on a Li-liquid target, we saw enormous rate enhancement in the $\mathrm{^7Li(d,n)^8Be\rightarrow2\cdot}$$\mathrm{ ^4He}$ reaction [7]. This enhancement (around $10^{10}$) was predicted by taking into account the fusion proceeded through coupling with a spontaneous chemical, i.e., atomic reaction forming an intermediate united atom (quasi-Be atom) where twin oscillatory nuclei coexist at the center of the common 1s-electron orbital. The spontaneous (coupled nuclear) reaction rate in the liquid is enhanced at a rate of $K\mathrm{= exp(-\Delta \textit{G}_r / \textit{k}_B \textit{T}})$ specified by the chemical potential (Gibbs free energy) change $\Delta G_{\mathrm{r}}$ of the reaction. Here $\mathrm{\textit{k}_B}$ denotes the Boltzmann constant, and $T$ the liquid temperature.

In this new scheme of enhanced pycnonuclear fusion in a Li liquid, we expect doubly intensified enhancement $K(\mathrm{Be}_2)=K^2(\mathrm{Be})$ with $\mathrm{\Delta \textit{G}_{r}(Be_2)= 2\Delta \textit{G}_r(Be)}$ by forming intermediate quasi-$\mathrm{Be_2}$ molecules under the implantation of slow molecular $\mathrm{D_2}$ ions [8]. An experiment comparing atomic- and molecular-ion implantation on the Li-liquid target supported this prediction [9]. The rate enhancement was close to $10^{20}$ as predicted [10] in the $\mathrm{^7Li(^7Li,2n)^{12}C}$ reaction [11]. As such, we surmise that the enhanced pycnonuclear reaction takes place in metallic Li liquids and condensed plasma in supernova progenitors. We study the enhancement of nuclear reactions in such metallic liquids as metallic hydrogen and Li liquids, taking into account the special role of s-electrons (i.e., they interpenetrate the nuclei and come into contact interaction with the internal nucleons). These interpenetrating s-electrons have a decisive influence on nuclear kinetics. We see this nuclear penetration effect of s-electrons in 1) $\beta$-decays through the electron capture, and 2) electric monopole ($E0$) isomeric transitions through the internal s-electron conversion [12].

The bulk of valence atomic electrons are thermodynamical. This results in astronomical rate enhancement in spontaneous chemical reactions in solutions. This thermodynamical liquid activity transforms contagiously into nuclear reactions with the collective dynamics of highly correlated itinerant s-electrons in metallic liquids or condensed plasma towards the chemical potential minimum [13]. In essence, this enhanced reaction is a united spontaneous chemical and nuclear reaction in metallic liquids or condensed plasma. In metallic liquids and condensed plasma in stars, the number density of conduction or itinerant s-electrons ranges from $10^{22}$ to $10^{23}$ cm$^{-3}$ for such light metals as Li, Na, and K. In these density regimes, effects arising from the exchange and Coulomb coupling between the s-electrons are called strongly coupled [14]. Itinerant s-electrons in metallic liquids can be regarded as Fermi liquids yielding thermodynamical liquid activity. Molecular ion implantation experiments on metallic Li liquids secured D-Li and $\mathrm{D_2}$-2Li fusion reactions enhanced with factors over 10 figures [10]. Evidence for the coherent formation of $\mathrm{^8Be^*_2}$ (i.e., a compound $\mathrm{^8Be}$ nucleus) in the $\mathrm{D_2}$-2Li reaction [15] and later corresponding heavy breakthrough phenomena was also obtained [9]. 

Ionic reactions in aqueous solutions are spontaneous reactions that take place with enormously large rate enhancement [16]. As such, 
\begin{align*}
\mathrm{Hg^{++} + S^{--} \leftrightarrow HgS(red)\,} 
\end{align*}
\begin{equation}
K = \overrightarrow{k} / \overleftarrow{k} = 2 \times 10^{53} \: \: \mathrm{at \: 25^{\circ}C}
\end{equation}
where $K$, $\overrightarrow{k}$ and $\overleftarrow{k}$ denote the equilibrium constant, the forward- and backward-reaction rates, respectively. Equation (1) means that the forward reaction rate $\overrightarrow{k}$ is enhanced, that is
\begin{equation}
\overrightarrow{k} = k_0 A \,\,
\end{equation}
where $k_0$ and $A$ denote the enhancement factor and the intrinsic reaction rate between two reactant ions, respectively. The backward reaction rate $\mathrm{\overleftarrow{k}}$ is reduced by $A^{-1}$, hence
\begin{align*}
K = k_{0}A/(k_0/A) = A^2
\end{align*}
\begin{equation}
A=K^{1/2}
\end{equation}

\hyphenchar\font=-1
Note that the ionic reaction specified by Eq. (1) takes place also in an alcohol solution with the same enhancement. In both water and alcohol, $\mathrm{Hg^{++}}$ and $\mathrm{S^{--}}$  ions undergo this ionic reaction with the same enhancement. Even without any solvent, this ionic reaction takes place in fused salts. In sum, this ionic reaction takes place with large enhancement, regardless of the kind of solvent.

This phenomenon is explained with Widom’s thermodynamical activity of liquids [13], which highlights the common bulk/collective features of the spontaneous reactions caused by thermodynamic force in those liquids dissolving reactant particles. The macroscopically distinct parts of liquids around reactant particles are correlated, hence a long-range coherence. The thermodynamic force is specified by the chemical potential change $\Delta \textit{G}_r$ in the reaction. This relationship is strictly independent of the kinds of reactant particles and the nature of the microscopic interparticle interactions [17].

We conducted ion implantation experiments on the surface of a metallic Li liquid to investigate the reaction enhancement. We found unstable reaction rates that are peculiar to the chemically reactive surface of a metallic Li liquid. For instance, at a vacuum of $10^{-7}$ Torr, it was hard to keep a clean Li surface within one second during the slow D$_2$ ion implantation without the dissociation of the $\mathrm{D_2}$ ions on the Li surface provided that a simultaneous slag film sputtering clean-up treatment was given. As such, a stabilizer is necessary for such liquid Li metals as Ni or Pd nanocrystal powders. In the combined system of metallike hydrides and Li, a donor of itinerant electrons may lead to self-sustained, super enhanced H-H nuclear fusion and successive He burning chemonuclear reactions. The hydrogen ions are strongly screened by valence electrons and nearly localized itinerant electrons in hybridized states in the metallike hydrides [18]. This effect is subject to the short-range screening length $D_{\mathrm{s}}$. Within this range, low energy fusion reactions are effectively enhanced. The ions are confined in their respective bond spaces with a number density
\begin{equation}
n_{\mathrm{i}} = (4\pi /3)^{-1} a_{\mathrm{i}} ^{-3}
\end{equation}
where $a_{\mathrm{i}}$ denotes the Wigner--Seitz radius of the ions [19]. 

The screened nuclear fusion rate per number density is
\begin{equation}
R_{\mathrm{s}} = \lambda _{\mathrm{HH'}} n_{\mathrm{H}} n_{\mathrm{H'}} = \frac{2S(0)r_{\mathrm{HH'}}^* }{(1 + \delta _{\mathrm{HH'}}) \hbar} \cdot \sqrt{\frac{D_{\mathrm{s}}}{r_{\mathrm{HH'}}^*}} \mathrm{exp} \left[ -\pi \sqrt{\frac{D_{\mathrm{s}}}{r_{\mathrm{HH'}}^*}} \right] \cdot n_{\mathrm{H}} n_{\mathrm{H'}}\: 
\end{equation}
between the hydrogen ions at a low temperature [1]. Here $\lambda_{\mathrm{HH'}}$ denotes the fusion constant, $\delta_{\mathrm{HH'}} = 1$ for the same kind of hydrogen $\mathrm{H'=H}$ and 0 for a different kind of hydrogen $\mathrm{H'\neq H}$. The factor $S(0)$ refers to the reaction cross-section factor. The nuclear Bohr radius $r_{\mathrm{HH'}}^*$ is represented by the electron mass $m_{\mathrm{e}} = 0.511\,\mathrm{MeV}/c^2$ (where $c$ is the speed of light), the average nucleon mass $m\mathrm{_N = 931.5\,MeV}/c^2$, the reduced ion mass $\mu _{\mathrm{HH'}}$, and the Bohr radius $a_{\mathrm{B}}$ as follows:
\begin{equation}
r_{\mathrm{HH'}}^* = \frac{m_{\mathrm{e}}}{2\mu_{\mathrm{HH'}} } \cdot a_{\mathrm{B}} = 1.45\times 10^{-14} \frac{A_{\mathrm{H}} + A_{\mathrm{H'}}}{A_{\mathrm{H}} A_{\mathrm{H'}} } \: \mathrm{(m)} \: 
\end{equation}
provided that
\begin{align*}
\mu_{\mathrm{HH'}} = \frac{A_{\mathrm{H}} A_{\mathrm{H'}} m_{\mathrm{N}}}{A_{\mathrm{H}} + A_{\mathrm{H'}}} \: 
\end{align*}
where $A_{\mathrm{H}}$ and $A_{\mathrm{H'}}$ denote the mass numbers of H- and H'- ions, respectively. 

\vspace{0.25cm}
\begin{center}
\small
\noindent Table 1.1 \, Cross-section factors $S(0)$ and Q-values
 \\of hydrogen burning fusion reactions.
\vspace{0.25 cm}
\small
\begin{tabular}{| c | c | c |}
\multicolumn{3}{ p{7cm} }{} \\ \hline
 &  $S(0)$ $\mathrm{(MeV\cdot b)}$  & $Q$ (MeV) \\ \hline
$\mathrm{H(p, e^+ \nu_e)D}$ & $3.4\times 10^{-25}$ & $1.4$ \\ \hline
$\mathrm{D(p,\gamma )^3He}$ & $2.5\times10^{-7}$ & $5.5$ \\ \hline
$\mathrm{D(d,p)T}$ & $0.053$ & $4.0$ \\ \hline
$\mathrm{D(d,n)^3He}$ & $0.050$ & $3.3$ \\  \hline
$\mathrm{D(d,\gamma)^4He}$ & $\sim 10^{-7}$ & $23.8$ \\ \hline
\end{tabular}
\end{center}

Table 1.1 shows the $S(0)$ and $Q$-values of the hydrogen burning fusion reactions [20, 21]. Equation (5) is thus written as follows:
\begin{equation}
\mathrm{H(p,e^+ \nu_{\mathrm{e}})D} \: : \; R_{\mathrm{s}} = 1.5\times 10^{-45} n_{\mathrm{H}}^2 \sqrt{\frac{D_{\mathrm{s}}}{2.9\times 10^{-14}}} \mathrm{exp} \left[ -\pi \sqrt{\frac{D_{\mathrm{s}}}{2.9\times 10^{-14}}} \right] \: 
\end{equation}
\begin{equation}
\mathrm{D(p, \gamma )^3He} \: : \; R_{\mathrm{s}} = 1.2\times 10^{-25} n_{\mathrm{H}} n_{\mathrm{D}} \sqrt{\frac{D_{\mathrm{s}}}{2.3\times 10^{-14}}} \mathrm{exp} \left[ -\pi \sqrt{\frac{D_{\mathrm{s}}}{2.3\times 10^{-14}}} \right] \: 
\end{equation}

\begin{equation}
\mathrm{D(d, p)T} \: : \; R_{\mathrm{s}} = 1.2\times 10^{-22} n_{\mathrm{D}}^2 \sqrt{\frac{D_{\mathrm{s}}}{1.5\times 10^{-14}}} \mathrm{exp} \left[ -\pi \sqrt{\frac{D_{\mathrm{s}}}{1.5\times 10^{-14}}} \right] \: 
\end{equation}

\begin{equation}
\mathrm{D(d, n)^3He} \: : \; R_{\mathrm{s}} = 1.1\times 10^{-22} n_{\mathrm{D}}^2 \sqrt{\frac{D_{\mathrm{s}}}{1.5\times 10^{-14}}} \mathrm{exp} \left[ -\pi \sqrt{\frac{D_{\mathrm{s}}}{1.5\times 10^{-14}}} \right] \: 
\end{equation}

\begin{equation}
\mathrm{D(d, \gamma)^4He} \: : \; R_{\mathrm{s}} \approx 10^{-28} n_{\mathrm{D}}^2 \sqrt{\frac{D_{\mathrm{s}}}{1.5\times 10^{-14}}} \mathrm{exp} \left[ -\pi \sqrt{\frac{D_{\mathrm{s}}}{1.5\times 10^{-14}}} \right] \: 
\end{equation}

\noindent where $R_{\mathrm{s}}$, $n$ and $D_{\mathrm{s}}$ are given in the units of $\mathrm{s^{-1} m^{-3}}$, $\mathrm{m^{-3}}$ and m, respectively.

In the 1970s, the possible formation of metallic hydrogen in certain transition metals was noticed during the systematic investigation of alloys [22]. The formation of transition metal hydride takes place in chemisorption in stages: hydrogen molecules dissociate on the surface of adsorbent Pd or Ni nanoclusters and are transformed into metallic hydrogen, and then the metallic hydrogen is alloyed. In this chapter, transition metal hydrides revealing metallic features are referred to hereafter as metallike hydrides. The electronegativity (Pauling Scale) of hydrogen $\chi$=2.2 is close to those of Ni ($\chi$=1.98) and Pd ($\chi$=2.2) [16]. The electronegativity of Li is $\chi$=0.98, and Ca $\chi$=1. When the Li and Ca s-electrons are transferred to the Pd and Ni cells to form an alloy, the boundary mismatch between Li/Ca and Ni/Pd metals is not relevant, because the Li and Ca atoms are a lot smaller than the cells. Li and Ca supply itinerant electrons to the Ni and Pd metals, lowering the melting point of adsorbed metallic hydrogen and triggering the enhanced cold nuclear fusion of the metallic hydrogen liquid. The same thing is expected for Na, K, Cs, Sr, and Ba. In the presence of few itinerant electrons, the metallic hydrogen forms a quantum solid rather than a semiclassical liquid in metallike hydrides (e.g., deuterons that perform zero-point oscillation around their lattice sites). The fusion rates are determined by Eqs. (9)--(11) based on the contact probabilities between adjacent nuclei.

The enhancement occurs with thermodynamical liquid activity, if at least one of the reacting nuclear species is in a liquid state. A melting temperature ${T_m}$ for hydrogen D derived from a Lindemann-type criterion [14] with the Thomas--Fermi approximation is
\begin{align*}
3k_{\mathrm{B}}T< \hbar^2 \lbrack \frac{(3\pi n_{\mathrm{e}})^ \frac{2}{3}}{2m}\rbrack
\end{align*}
\begin{equation}
T_{\mathrm{m}}= (\frac{e^2}{180a_{\mathrm{D}}k_{\mathrm{B}}})\{1+(\frac{20}{3\pi})(\frac{\alpha}{\lambdabar_{\mathrm{e}}})(\frac{n_{\mathrm{e}}}{12n_{\mathrm{D}}^2})^\frac{1}{3}\}^{-1} 
\end{equation}

\noindent
where ${a_D}$ denotes the Wigner-Seitz radius of the deuteriums, ${\lambdabar_e}$ the Compton wavelength of the electrons, ${n_e}$ the number density of the electrons, $m$ the electron mass, and $\alpha$ the fine structure constant. 

Equation (12) gives the intrinsic melting temperature of metallic deuterium in metallike hydrides with few density itinerant electrons. In PdD of $\mathrm{n_D}$ $\approx$ $6.3\times 10^{28} m^{-3}$ (see Table 1.2 and [14]), the intrinsic melting temperature is ${T_m}$ = ${e^2}/({180a_{\mathrm{D}}k_{\mathrm{B}}})$ $\approx$ 300 K, suggesting that the liquid activity of metallic deuterium is critical at room temperature. However, in metallike hydrides containing itinerant electrons of ${{n_e}}$ $\approx$ $10^{26} m^{-3}$, the melting temperature of the metallic deuterium drops to around 200 K and the metallic deuterium reveals thermodynamical liquid activity at room temperature. As a result, the factor of fusion rate enhancement exceeds $10^{30}$. 
\vspace{0.25cm}
\begin{center}
\small
\noindent Table 1.2 \,Parameters of Eqs. (7)--(11) for screened low-temperature nuclear fusion rate and power released in metal hydrogen systems.
\vspace{0.25 cm}
\small
\begin{tabular}{| c | c | c | c |}
\multicolumn{4}{ p{9.6cm} }{} \\ \hline
 & $\mathrm{TiH_2/TiD_2}$ & NiH/NiD & PdH/PdD \\ \hline
$n\mathrm{_H / }n\mathrm{_D (m^{-3})}$ & $9.4\times 10^{28}$ & $7.3\times 10^{28}$ & $6.3\times 10^{28}$ \\ \hline
$D\mathrm{_s (10^{-11}m)}$ & 2.8 & 2.2 & 1.9 \\ \hline
$R_s\mathrm{(s^{-1} \cdot m^{-3}), H-H}$ & $1.7\times 10^{-28}$ & $5.8\times 10^{-24}$ & $1.8\times 10^{-21}$ \\ \hline
$R_s\mathrm{(s^{-1} \cdot m^{-3}), D-D}$ & $1.0\times 10^{-21}$ & $2.6\times 10^{-15}$ & $8.9\times 10^{-12}$ \\ \hline
$P\mathrm{(W \cdot m^{-3}), H-H}$ & $1.5 \times 10^{-41}$ & $5.2\times 10^{-37}$ & $1.7\times 10^{-34}$ \\ \hline
$P\mathrm{(W \cdot m^{-3}), D-D}$ & $5.9\times 10^{-34}$ & $1.5\times 10^{-27}$ & $5.2\times 10^{-24}$ \\ \hline
\end{tabular}
\end{center}

In experiments using the electrochemical method, a nanostructure Pd cathode and an alkali-${\mathrm{D_2}}$O solution (e.g., Cs${\mathrm{NO_3}}-{\mathrm{D_2}}$O solution) are necessary. On the surface of Pd cathode, ${\mathrm{D_2}}$O is reduced, thereby producing deuterium that undergoes chemisorption forming PdD. At the same time, Cs${\mathrm{NO_3}}$ is reduced but most of the Cs atoms produced immediately react with the solution forming CsOD. However, few of these Cs atoms are taken into Pd cathode as a part of the electron donor. Ordinary itinerant electrons are slim. Their action is critical to reducing the melting temperature of the adsorbed metallic deuterium. This applies to the LiOD-${\mathrm{D_2}}$O solution electrochemical method. We therefore introduce metallike hydrides--electron donor mixtures by combining Ni nanocrystal powder with LiD or its complex compound (e.g., LiAl${\mathrm{D_4}}$ or LiMg${\mathrm{D_3}}$) in an atmosphere of ${\mathrm{D_2}}$ gas. 

\hyphenchar\font=-1
\section*{1.3 Elements of the Chemonuclear Fusion}
\noindent
In the fusion system of metallike hydride--electron donor mixtures, the dimensionless de Broglie wave length $\Lambda _{\mathrm{i}}$ of hydrogen atoms with mass number $A\mathrm{_i}$ is very small, that is
\begin{equation}
\Lambda_{\mathrm{i}} = \frac{\hbar}{a_{\mathrm{H}}} \left( \frac{2\pi}{A_{\mathrm{i}} m_{\mathrm{N}} k_{\mathrm{B}} T} \right)^{\frac{1}{2}} < 1 \: 
\end{equation}
where $a_{\mathrm{H}}$ denotes the Wigner-Seitz radius in Eq. (4). 

The wave mechanical effects on the atoms are therefore negligible. Hydrogen pairs reacting within their bonds do not disturb much the surrounding spectator atoms in the molecules, because the interaction can be treated as one between two screened particles. These features confirm the validity of rate enhancement evaluation in the scheme of semiclassical dynamics, as far as nuclear fusion under the hydrogen ion conditions is concerned. Their dynamics can be described as electron screened ion cores immersed in the sea of itinerant electrons. 

Within the bonds, the hydrogen atoms collide with each other (e.g., at $T$ = 773 K) with the frequency
\begin{equation}
\nu_{\mathrm{H-H}} = k_{\mathrm{B}} T / \hbar = 1\times 10^{14} \: \mathrm{s^{-1} }\: 
\end{equation}
that is a lot smaller than the gyration frequency $\nu_{\mathrm{s}}$ of the 1s-orbital electrons of hydrogen atoms. The latter is
\begin{equation}
\nu_{\mathrm{s}} = \upsilon_{\mathrm{B}} / 2\pi a_{\mathrm{B}} = \alpha c / 2\pi a_{\mathrm{B}} = 7\times 10^{15} \: \mathrm{s^{-1}} \: 
\end{equation}
where $\upsilon_{\mathrm{B}}$ is the Bohr speed, $\alpha$ is the fine structure constant, and $c$ is the speed of light. The orbital electrons adjust their electronic states continuously and smoothly to the nuclear collision process. Nuclear fusion takes place when the hydrogen bonds collapse. As a result, intermediate united atoms (quasi-He atoms) with pairs of oscillatory hydrogen nuclei at the centers of the common 1s-orbitals come into being. In ion-ion colliding experiments, these united atoms are often detected through x-ray spectroscopic measurement.

During the collapse, the volume of two screened hydrogen atoms shrinks by one order of magnitude in association with the chemical potential change $\Delta G_{\mathrm{r}}$ [10]. Normally most of the quasi-He atoms do not necessarily proceed to the nuclear fusion and decay into pairs of hydrogen atoms. In the chemonuclear fusion system, the hydrogen bonds are metallic and liquefied in the sea of itinerant s-electrons. This sustains the shrinking process and prolongs the lifetimes of the quasi-He atoms with a factor of $A= K^{1/2} = \mathrm{exp}\left[ -\Delta G_{\mathrm{r}} / 2k_{\mathrm{0}} T \right]$ based on Eq. (3), resulting in the enormously enhanced rate $KR_s$ of the chemonuclear fusion, thanks to the zero-point oscillation expressed in Eqs. (7)--(11).

These results are based on the thermodynamical conception that a fractional change of the reaction rate is exactly proportional to an entropy change in the universe [17]. In addition, these results are based on the chemical potential (Gibbs free energy) and the chemical potential change $\Delta G_r$ defined at the normal electron density on Earth $n_e=0.03$ atomic units (a.u.) $= 0.03({4\pi}/{3})^{-1}a_\mathrm{B}^{-3}$ in particular. The itinerant s-electron density of such metallike hydride and electron donor as Li mixtures, however, is highly dependent on the kinds of electron donors and their mixing amount (e.g., $n_e = 0.085$ a.u.).

The chemical potential is proportional to $n_e^{2/3}$, hence the chemical potential change in the mixtures

\begin{align*}
\Delta G_r \mathrm{(normal}\,n_{e^-}) \longrightarrow \Delta G_r(n_{e^-}=0.085\, \mathrm{a.u.})
\end{align*}
\begin{align*}
= (0.085/0.03)^{2/3}\Delta G_r(\mathrm{normal}\,n_{e^-})=2\Delta G_r \, 
\end{align*}
\noindent
We therefore use the equilibrium constant $K$ as the enhancement $A$ in the dense electron donor mixtures, hence
\begin{equation}
A=\mathrm{exp}\,[-2\Delta G_r/2k_\mathrm{B}T]=\mathrm{exp}\,[-\Delta G_r/k_\mathrm{B}T]=K\,
\end{equation}
The above equation describes chemonuclear reaction enhancement in the system of metallike hydride--electron donor mixtures, where $\Delta G_r$ denotes $\Delta G_r$ (normal $n_{e^-}$) for brevity unless otherwise specified.

The nuclear fusion rate enhancement $(\Delta G_\mathrm{r} < 0)$ was derived by astroplasma physicists based on quantum statistical mechanics [1, 3, 4]. However, Bohr, who developed the concepts of \textit{nucleus} and \textit{Rutherford scattering} [23], argued that the latter was not simply a collision between nuclei but a collision between whole atoms at low energies, especially several keV/amu [24]. This phenomenon is called the nuclear stopping of ions/atoms moving in a condensed matter, because the nuclei of struck ions/atoms acquire an enormous amount of kinetic energy in the collision [25]. We infer that collisions between nuclei are fully linked to atomic collisions which are strictly governed by statistical thermodynamics. This idea is key to the chemonuclear reaction in which nuclear reactions take place together with the atomic processes and thus are controllable through the synchronous formation of atomic or nuclear intermediate complexes under the right chemical or physical conditions [6-12]. We take the D-D atomic collision as an example. This collision shows the Rutherford scattering between two deuterons dressed with atomic electrons. They are directed to the nuclear collision at their screening distance $\mathit{D_s}$ (e.g., 19 pm) in a Pd nanocluster powder (\textit{cf.} Table 1.2). The colliding D-atoms will form a united atom (quasi-He atom) in which twin deuterons coexist and are confined at the center of the common quasi-He K-shell electron orbital of radius 26 pm. In the quasi-He atom, twin deuterons collide with zero-point oscillation energy towards the pycnonuclear fusion, thanks to the tunnelling effect.

The united atom (quasi-He atom) corresponds to the intermediate complex that decays into the final product in the reaction. We apply the rate equation of chemical reaction through intermediate complex formation to the united atom formation. We therefore reformulate the Arrhenius equation as 
\begin{align*}
\overrightarrow{k} = k_0 \,\mathrm{exp}[-\Delta \textit{G}_r / \textit{k}_BT] = k_0K \tag{16'}
\end{align*}
in the scheme of Gibbs statistical thermodynamics. The chemical potential change $\Delta G_r$ in the reaction corresponds to the activation energy $E_a$ in the Arrhenius equation. In the reversible reaction, $\Delta G_r > 0$, while $\Delta G_r < 0$ in the spontaneous reaction revealing the reaction enhancement $K > 1$.

For the fusion reaction producing a compound He nucleus via a quasi-He atom formation
\begin{equation}
\mathrm{D + D \longrightarrow quasi\mbox{-}He \longrightarrow He} \: 
\end{equation}
the chemical potential change $\Delta G_{\mathrm{r}}$ is evaluated by
\begin{equation}
-\Delta G_{\mathrm{r}} (\mathrm{quasi\mbox{-}He} ) = -\Delta G_{\mathrm{r}} (\mathrm{He}) = \phi^* (\mathrm{He}) - 2 \phi^* (\mathrm{D}) 
\end{equation}
The change of the chemical potential $\phi^*$ of impurity atoms in bulk metals [22] is approximated by the change of immersion energy $\Delta E^{hom}$ of atoms embedded in an electron gas/liquid. The immersion energy has been evaluated as a function of the electron density in atomic units ${3}/({4\pi{{a_B}}^3})=1.6\times10^{30} \mathrm{m}^{-3}$. 

At such low density as ${{n_e}} < 0.01\,a.u.$, $\Delta E^{hom}$ is negative with a minimum point for the atoms with stable free negative ions such as H$^-$ ions [26]. 
We evaluate the plausible value of $\phi^*$(He) using Eq. (19) derived from the alloy data on $\phi^*$, i.e., Li: 2.85 eV, Be: 4.2 eV, B: 4.8 eV, C: 6.23 eV, N: 7 eV [32]. Firstly,
\begin{equation}
\phi^* = 1.25(Z-1)+0.03\:\textrm{eV}
\end{equation}
\begin{equation}
\phi^*\textrm{(He)} = 1.28\:\textrm{eV}
\end{equation}
Secondly,
\begin{equation}
\phi^*\textrm{(D)} = \phi^*\textrm{(H)} = 0.03\:\textrm{eV}
\end{equation}
Then we have
\begin{equation}
-\Delta G_{\mathrm{r}}\textrm{(He)} = 1.22\:\textrm{eV}
\end{equation}

Recalling Eq. (16), we express the enhancement factor of the chemonuclear D-D fusion via the collapse of D-D bond immersed in the fusion system as
\begin{equation}
K(\mathrm{He}) = 2.9\times 10^{20} \: \: \mathrm{at} \: T = 300\:\mathrm{K} \: (k_{\mathrm{B}} T = 0.026\:\mathrm{eV})
\end{equation}
The enhancement is reduced at higher temperatures, for instance, 
\begin{equation}
K(\mathrm{He}) = 2.4\times 10^{13} \: \: \mathrm{at} \:  T=460\:K
\end{equation}

\noindent This treatment may result in underestimated chemical potential in the system of metallike hydride-electron donor mixtures and Ni hydride-Li mixtures in particular [22]. We apply another formula derived from alloy data to metallic elements Li and Be, instead of Eqs. (19)--(24), hence
\vspace{-0.1cm}
\begin{align*}
\phi^* = 1.4(\mathrm{Z}-1)\:\mathrm{eV} \tag{19'}
\end{align*}
\begin{align*}
\phi^*(\mathrm{He}) = 1.4\:\mathrm{eV}  \tag{20'}
\end{align*}
\begin{align*}
\phi^*(\mathrm{D})=\phi^*(\mathrm{H})=0 \tag{21'}
\end{align*}
\begin{align*}
-\Delta G_\mathrm{r}\mathrm{(He)} = 1.4\:\mathrm{eV} \tag{22'}
\end{align*}
\begin{align*}
K\mathrm{(He)} = 1.4\times10^{25}\,\mathrm{at\,T=300\:K} \tag{23'}
\end{align*}
\begin{align*} 
K\mathrm{(He)} = 2.2\times10^{15}\,\mathrm{at\,T=460\:K} \tag{24'}
\end{align*}

As deuterium ion clusters undergo the fusion, the rate enhancement is intensified by coherent bond collapse [8]. A coherent chemonuclear fusion of molecular $\mathrm{D_2}$ ions implanted with a slow speed $v_{\mathrm{i}} < \alpha Z_{\mathrm{Li}} c$ on the surface of a metallic Li liquid was observed [9]. During the collision, pairs of atoms in the ions keep their correlation, inducing coherent atomic/ionic collisions. This phenomenon became known in the 1970s (e.g., the coherent sputtering phenomena of molecular ions) [27]. In addition, the rate enhancement of the molecular $\mathrm{D_2}$ ion-induced fusion was close to the square of the rate enhancement of the deuteron-induced fusion [9]. The coherent production of twin $\alpha$-particle pairs in the molecular $\mathrm{D_2}$ ion-induced fusion shows the coherent decay of an intermediate compound nuclear pair $\mathrm{(^8 Be)_2}$ [15]. This is similar to the coherent decay of positronium molecule $\mathrm{Ps_2}$ into twin-degenerate annihilation photon pairs [28].

\hyphenchar\font=-1
\section*{1.4 D-D Chemonuclear Fusion}
\noindent
For the coherent $\mathrm{D_n}$-$\mathrm{D_n}$ chemonuclear fusion
\begin{equation}
\mathrm{D_n + D_n} \longrightarrow (\mathrm{quasi\mbox{-} He})_n\longrightarrow (\mathrm{He})_n \: 
\end{equation}
the chemical potential change $\Delta G_{\mathrm{r}} (\mathrm{He_n})$ is given by Eqs. (22) and (22') as follows:
\begin{equation}
\Delta G_{\mathrm{r}} (\mathrm{He_n}) = n\Delta G_{\mathrm{r}} (\mathrm{He}) = -1.22n\:\mathrm{eV} \sim -1.4n\:\mathrm{eV}
\end{equation}

Note that He$_n$ molecules consisting of excited He atoms are stable. The enhancement factors of the chemonuclear fusion at $T=460\:K$ via the $n$-fold, doubly and trebly coherent collapse of D-D bonds are
\begin{equation}
K(\mathrm{He_n}) = K^n (\mathrm{He}) = (2.4\times 10^{13})^n \: 
\end{equation}
\begin{equation}
K(\mathrm{He}_{2}) = 5.7\times 10^{26} \: , \: K(\mathrm{He}_{3}) = 1.4\times 10^{40}
\end{equation}

For the system of Ni hydride-Li mixtures, we have 
\begin{align*}
K(\mathrm{He_n}) = K^n (\mathrm{He}) = (2.2\times 10^{15})^n \:  \tag{27'}
\end{align*}
\begin{align*}
K(\mathrm{He}_{2}) = 5.1\times 10^{30} \: , \: K(\mathrm{He}_{3}) = 1.2\times 10^{46} \tag{28'}
\end{align*}

In the mixed H and D systems, however, the coherent collapse of two H-D bonds is unlikely, because other coherent H-H or D-D collapse is more favorable. The $\mathrm{D(p,\gamma)^3He}$ reactions via coherent H-D bond collapse are therefore not realized.

In the Li permeated transition metal/hydrogen systems without deuteride or $\mathrm{D_2}$ gas, the D-D chemonuclear fusion still takes place through the coherent $\mathrm{H_2}$-$\mathrm{H_2}$ chemonuclear fusion. In the $\mathrm{H(p,e^+\nu_e)D}$ reactions via the collapse of H-H bonds, D atoms/ions are produced at rest, because their recoil energy in the positron and neutrino emission is below 0.1 eV. In this case, new D-D bonds are formed through the coherent $\mathrm{H_2}$-$\mathrm{H_2}$  chemonuclear fusion.

As the diffusion coefficient reaches $D_{\mathrm{o}} = 4\times 10^{-8} \: \mathrm{m^2 /s}$ in the $\mathrm{Mg_2 NiD_4}$ crystal, the D atoms/ions diffuse in the metal hydrogen system [29]. The D atoms that have been produced tend to cluster at the same sites in the crystal due to the Bose--Einstein condensation, hence the coherent $\mathrm{D_n}$-$\mathrm{D_n}$ chemonuclear fusion. Around ten percent of the hydrogen ions from the conventional PIG-type ion sources are molecular ions H$_2^+/$D$_2^+$ and H$_3^+/$D$_3^+.$ Given that the intrinsic H-H and D-D nuclear fusion rates differ by a factor of billions in the Ni- and Pd-hydrides, we device a control scheme of energy released in the fusion by tuning precisely the fraction of $\mathrm{D_2}$ gas. See Table 1.2 for the parameters pertinent to the electron screened nuclear fusion reactions [14]. The values for NiD have been estimated from those for Ti$\mathrm{D_2}$ and PdD based on the data showing that both the heat of hydrogen dissolution and lattice parameter of Ni are more close to those of Pd than those of Ti [29, 30]. The $\mathrm{D(p,\gamma)^3 He}$ reaction has been neglected, because a significant contribution from coherent bond collapse is unlikely in the H-D chemonuclear fusion.

As for Pd and Ni (Table 1.2), the power released in the D-D fusion is far stronger (by a factor over one billion) than that in the H-H fusion. We infer that the D-D chemonuclear fusion still takes place in a system of Li permeated metal and pure $\mathrm{H_2}$. The rate of chemonuclear fusion shows clearly that almost all energy released in the pure hydrogen system is from the D-D fusion in succession of the coherent H-H fusion and the D-D fusion caused by D$_2$ gas (which is around 0.017\% of ordinary hydrogen gas). 

The thermodynamical activity of liquids, as seen in supernova progenitors, can be reproduced in the combined system of electron liquids and metallike hydrides. Certain transition metal hydrides are metallike in that their hydrogen atoms are partly in the metallic state [22, 31, 32]. In the presence of dense itinerant electrons, around ten percent of such hydrogen atoms are likely to reveal thermodynamical liquid activity. Hydrogen adsorbing Ni or Pd nanocluster powders or nanostructured foils combined with Li or Na metals or alkaline earth oxides CaO or SrO can be the donors of dense electrons. Specifically, the Li metal may become the source of itinerant ions and electrons. The Li metal can be replaced with Li hydride LiH(D) or its complex (e.g., LiAlH$_4$(D$_4$)), because the hydride molecules dissociate on the surface of nanoclusters and the liberated hydrogen atoms are adsorbed into them. The Li hydride bond strength is weaker than the combined bond strength of NiH(D) and NiLi. Following this prescription [34], the experimental results have been improved by increasing the amount of Li and $\mathrm{D_2}$ [35, 36].

In the coherent $\mathrm{D_2}$-$\mathrm{D_2}$ and $\mathrm{D_3}$-$\mathrm{D_3}$ chemonuclear fusion reactions, the rates forming $\mathrm{He_2}$ and $\mathrm{He_3}$ complexes are respectively
\begin{equation}
R(\textrm{D}_2-\textrm{D}_2) = 3.3 \times 10^{-28} n^2_{D_2} \cdot  fK^2
\end{equation}
\begin{equation}
R(\textrm{D}_3-\textrm{D}_3) = 3.3 \times 10^{-28} n^2_{D_3} \cdot  fK^3
\end{equation}
where $f$ denotes the penetration factor
\begin{equation}
f = \sqrt{\frac{D_S}{1.5\times10^{-14}}} \: \mathrm{exp} \:  \Bigg[ -\pi \sqrt{\frac{D_S}{1.5\times10^{-14}}\Bigg]}
\end{equation}
and $K$ = $2.4\times10^{13}$ at $T$=460 K as in Eq. (24) or $K$ = $2.2\times10^{15}$ at $T$=460 K as in Eq. (24').  

The key feature of this coherent fusion is the radiationless decay of the intermediate fusion complexes quasi-$\mathrm{He_2}$ and quasi-$\mathrm{He_3}$. These complexes decay into double and triple He ions with kinetic energy of 23.8 MeV/He ion without causing any recoil effect. All these He ions induce successive reactions, making element synthesis possible. This finding gives us an avenue to achieve radioactive waste vanishing and the mass synthesis of noble metals.

\begin{figure}[h!]
\centering
\includegraphics[scale=0.2]{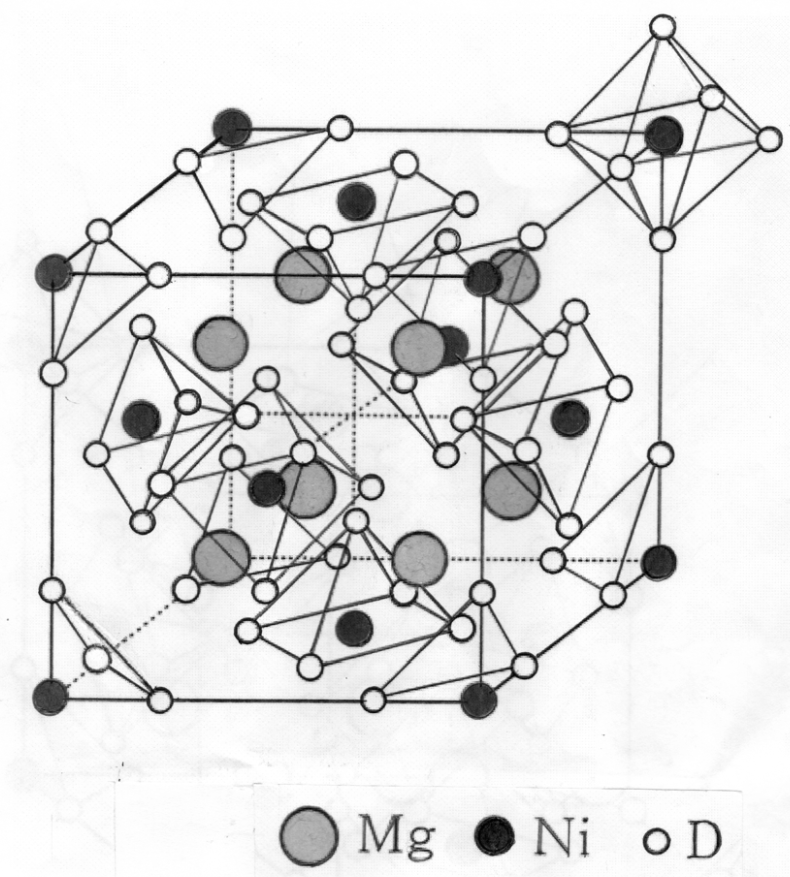}
\end{figure}
\small
Fig. 1.1 \, Crystal structure of $\mathrm{Mg_2 NiD_4}$ molecules. 
\vspace{0.5cm}
\normalsize

We give the treatment for the hydride NiH(D) to a fusion test system of $\mathrm{Mg_2 Ni H_4}$/$\mathrm{Mg_2 Ni D_4}$. The crystal structure of $\mathrm{Mg_2 NiD_4}$ [18, 29, 31] is shown in Fig 1.1. The hydrogen (H/D) ions cluster up to six at octahedral sites, making the trebly coherent $\mathrm{D_3}$-$\mathrm{D_3}$ chemonuclear fusion possible. The hydrogen (H/D) ions are strongly screened by and correlated with valence electrons and nearly localized itinerary s-electrons in a hybridized state in molecules immersed in the metallic Li liquid. These effects are specified by the short-range screening length $\mathrm{D_s}$ and the chemical potential change $\mathrm{\Delta G_r}$ in the fusion respectively. The enormously enhanced rate of chemonuclear fusion passes through the coherent collapse of hydrogen bonds.

We charge a test system of $\mathrm{Ni/Mg_2 Ni}$ nanocrystal powder mixed with $\mathrm{LiH/LiAlH_4}$ powder and hydrogen gas. The Li atoms play the role of the electron donor. With a 10-bar pressure on the gas, the chemical potential rises by 9kJ/mol=0.09 eV/molecule at $T$=773 K ($k_{\mathrm{B}}T$=0.067 eV) [37], hence a gain in hydrogen dissolution (into the $\mathrm{Ni/Mg_2Ni}$ crystals) with a factor of exp(0.09/0.067) = 3.8. This means a gain in the fusion rate and power released by a factor of 15. As a result, we convert the $\mathrm{Ni/Mg_2 Ni}$ crystals substantially to the hydride $\mathrm{NiH/Mg_2 Ni H_4}$ crystals in which around ten percent of the hydrogen atoms occupy octahedral sites in the hydride crystals forming 3-pairs. The trebly coherent collapse of the $\mathrm{D_3}$-$\mathrm{D_3}$ bonds with the enhancement of $1.4\times 10^{40}$ $\sim 12\times10^{46}$ at 460 K takes place following Eq. (28) and Eq. (28').

We preheat this test system at room temperature. The $\mathrm{LiAlH_4}$ molecules dissociate and form the $\mathrm{LiH}$ and $\mathrm{AlH_3}$ molecules. With physi-chemisorption on the surface of $\mathrm{Ni/Mg_2 Ni}$ nanocrystals, the LiH molecules dissociate faster than the hydrogen molecules, thanks to their large dipole moment $p$(LiH)$=2\times 10^{-29}(\mathrm{C\cdot m})$ and weak bond strength $D$(Li-H)$=2.47\:\mathrm{eV}$. Such physi-chemisorption is also observed in the cold nuclear fusion caused by the electrolysis of $\mathrm{D_2}$O containing LiOD using Pd electrodes. The dissociated H atoms are adsorbed by the nanocrystals and transformed into metallic hydrogen with Li$^+$ ions and itinerant s-electrons, hence the thermodynamical activity of the metallic hydrogen liquid. 

When the Li ions and itinerant s-electrons moisten the hydride powder, the coherent H-H fusion starts up with a large enhancement and heats up the hydride quickly. This reaction produces D-atoms in the hydride together with lattice vacancies refilled with H-atoms in the presence of the pressure H$_2$ gas. The D-atoms diffuse with the coefficient of diffusion ${D_0 \sim 10^{-7} \: m^2/s}$ at high temperatures towards the full clustering at octahedral sites [38, 39], resulting in the trebly coherent $\mathrm{D_3}$-$\mathrm{D_3}$ fusion and the release of an enormous amount of energy.

The most striking feature of H-H fusion in Ni hydrides is the $\gamma$-ray missing positron annihilation. Most positron annihilation $\gamma$-rays produced in the H-H fusion undergo the electron conversion associated with nuclear charge shake-off Auger electron $\textrm{e}^-_A$ emission, that is
\begin{equation}
\gamma + \textrm{Ni} \longrightarrow \textrm{e}^- + \textrm{e}^-_A + \textrm{Ni}^{2+}
\end{equation}
With thermodynamical liquid activity, this process is enhanced as
\begin{equation}
K = \mathrm{exp} \: \lbrack \frac{-\Delta G_r(\textrm{Ni}^{2+}_{\textrm{liq}})}{k_BT}\rbrack = \mathrm{exp} \: \lbrack \frac{-\Delta G_f(\textrm{Ni}^{2+}_{\textrm{aq}})}{k_BT}\rbrack \approx 2\times10^5
\end{equation}
where $\Delta G_f$ is the formation Gibbs free energy at $T$=460 K with $-\Delta G_f(\textrm{Ni}^{2+}_{\textrm{aq}})$ =  0.48 eV [16]. We assume that $\textrm{Ni}^{2+}_{\textrm{liq}} = \textrm{Ni}^{2+}_{\textrm{aq}}$ in an aqueous solution. The conversion coefficient is around ten percent in magnitude. All annihilation $\gamma$-rays have been converted into electrons. These electrons generate numerous soft x-rays in successful cold fusion experiments [34, 35, 41]. 

We make the following predictions:
\begin{enumerate}[1.]
\item In a chemonuclear fusion test system charging Li permeated \\$\mathrm{Ni/Mg_2Ni}$ nanocluster powder and hydrogen gas, we expect the trebly coherent chemonuclear $\mathrm{H_3}$-$\mathrm{H_3}$ and $\mathrm{D_3}$-$\mathrm{D_3}$ fusion but not the H-D fusion which generates 5.5 MeV $\gamma$-ray.
\item In the system charging ordinary hydrogen gas, the contribution of the H-H fusion is 1.1\% of the total power released. Most power is released in the $\mathrm{D_2}$-$\mathrm{D_2}$ and $\mathrm{D_3}$-$\mathrm{D_3}$ fusion caused by D$_2$ gas. Even if we take into account the fewer adsorption efficiency of D$_2$ gas by around 8\% compared to H$_2$ gas [40], the contribution of the H-H fusion is merely 4.4\%.
\item As for the coherent fusion enhancement of around $10^{40}\sim10^{46}$ at $T$=460 K and ${n_{D_3}}/{n_D}\approx 10^{-2}$ in the Ni deuteride, we expect the power output of the $\mathrm{D_3}$-$\mathrm{D_3}$ fusion to be $\mathrm{MW \cdot m^{-3}\sim GW \cdot m^{-3}}$ which is far greater than the solar interior power density.
\item In the coherent $\mathrm{D_n}$-$\mathrm{D_n}$ fusion, the intermediate fusion complexes (quasi-He$_n$) undergo radiationless decay, yielding He ions of 23.8 MeV kinetic energy. These ions undergo successive He-induced chemonuclear reactions.
\end{enumerate}

We have seen many experiments that failed to achieve stable and self-sustained cold nuclear fusion. These experiments focused on the detection of 1) calorimetrical signals, 2) annihilation $\gamma$-rays following the H(p,$\textrm{e}^+$$\nu$)D reaction, and 3) neutrons produced in the D(d,n) $^3$He reaction. Insufficient itinerant electron density in the reaction space (e.g., metallike hydrides) causes the instability of cold nuclear fusion. The role of itinerant s-electrons in the chemonuclear fusion has gone unnoticed. In the absence of sufficient itinerant s-electron density, neither a decrease in the melting point of adsorbed metallic hydrogen nor thermodynamical liquid activity is likely. Thermodynamical liquid activity makes the annihilation $\gamma$-rays due to the enhanced electron conversion process in Eqs. (32) and (33) possible. In metallike hydride--electron donor mixtures, enhanced electron conversion takes place. Unless an appropriate ($\alpha$,n) reaction converter (e.g., a lump of Be) is placed in the mixtures, we may not observe neutrons associated with the cold nuclear fusion. The D(d,n)$^3$He reaction is suppressed in the presence of thermodynamical liquid activity that enormously enhances the neutronless coherent $\mathrm{D_n}+\mathrm{D_n}\longrightarrow n^{4}$He $(n > 1)$ reaction. In spite of the spectroscopically and chemically confirmed evidence of $^4$He produced in the D-D cold fusion [41], the reaction $\mathrm{D+D} \rightarrow \mathrm{^4He} + 23.8 \, \mathrm{MeV}$ is inhibited on the basis of two-body kinematics. The observation of $^4$He -- a major puzzle of cold nuclear fusion -- is in strong support of the chemonuclear fusion.

We deal with nuclear fusion reactions at moderate temperatures ranging from 460 to 770 K with the implicit assumption of temperature independent chemical potential. For the system charging Li of certain weight percent of Ni or Ni hydrides (NiD$_2$ or NiH$_2$), this implicit assumption is not adequate. With a sufficient amount of Li supplying itinerant s-electrons, the chemical potential rises dramatically as the temperature soars. The electron work function of Li is 1.4 eV. As such, the density of $n_e$ itinerant electrons supplied by Li increases quickly alongside the temperature (e.g., with a factor of higher orders of magnitude at 1200 K compared to that at 460 K). The chemical potential varies in proportion to $n_e^{2/3}$ except for extremely high-density electrons. In this system, the enhancement factor exp[$-\Delta G_\mathrm{r}/T$] does not decrease in magnitude. The reaction rate may increase to 1700 K in this system. At this temperature, the stability of the Ni nanocluster powder is no longer guaranteed [35, 36, 41].

\hyphenchar\font=-1
\section*{1.5 A Supernova on Earth}
\noindent
A widespread belief is that nuclear reactions are not influenced by the chemical reactions of atoms and molecules around the reactant nuclei. In metallic hydrogen and lithium liquids of collective itinerant s-electrons and nuclei, however, nuclear reactions are subject to the thermodynamical activity of liquids due to the coupling between the bulk of nuclei and s-electrons. The s-electron density with respect to the nuclear positions plays an essential role in chemonuclear fusion in metallic Li liquids. In those liquids, all extraordinary enhancement mechanisms are subject to the thermodynamic force specified by the chemical potential change of the reaction. 

In the presence of itinerant electrons in the metallike hydride, the melting point of adsorbed metallic hydrogen will decrease, hence the formation of the liquid system of hydrogen ions and itinerant electrons. This system reveals the thermodynamical activity of liquids, inducing the super enhanced chemonuclear fusion. At extremely high temperatures, the enhancement $K=\mathrm{exp}\: \lbrack {-\Delta G_r}/{k_BT}\rbrack$ will be reduced. We have to see if the chemonuclear fusion scheme is applicable to the pycnonuclear C-C and C-O fusion in the white dwarf progenitors of supernovae. Nuclear fusion rates in the dense binary mixtures of carbon and oxygen are essential quantities that govern the evolution and ignition of the white dwarf progenitors of supernovae [14]. A first-principles calculation of nuclear fusion rates in dense C-O binary ionic mixtures of equimolar fraction was done in the fluid phases [42]. Enhancement factors of $10^{24}$ and $10^{30}$ were reported for the C-C and C-O nuclear fusion reactions respectively at the mass density $4\times10^9 {g}/{cm^3}$ and the temperature $10^8$ K in a state of Fermi degeneracy for electrons ${k_BT}/{E_F}=2\times10^{-4}$.

We relate the chemical potential of those elements immersed in the sea of high-density electrons to their immersion energies [26]. The immersion energies $\Delta E^{hom}$ are associated with 1) the chemical potential effects due to the addition of $Z$ electrons on the Fermi level in embedding the atoms of atomic number $Z$, and 2) the relaxation effects due to improved screening in the metallic environment. Hence
\begin{equation}
\Delta E^{hom}(n_e) = Z\bar{\mu}(n_e) + \Delta E_R(n_e)
\end{equation}
where $\bar{\mu}$ is the internal chemical potential of the electron gas, and $\Delta E_R$ is the relaxation or rearrangement energy. At higher densities, the relaxation energy $\Delta E_R$ is only weakly dependent on density (i.e., extremely high-density limit $\Delta E_R(n_e)\propto n_e^{\frac{1}{6}}$) and it will compensate for a rapid increase $(\propto Zn_e^{\frac{2}{3}})$ in the chemical potential. 

This process suggests a dimension of $Z_{eff}$. The effective number of electrons in an atom that are sensitive to the environment is 
\begin{equation}
Z_{eff}=\frac{\frac{d\Delta E^{hom}}{dn_e}}{\frac{d\bar{\mu}}{dn_e}}
\end{equation}
The chemical potential $\phi^*(Z)$ of those elements with the atomic number Z immersed in the extremely high-density electrons is
\begin{equation}
\phi^*(Z) \sim Z_{eff}\bar{\mu}
\end{equation}
instead of $Z\bar{\mu}$, where
\begin{equation}
\bar{\mu} = 2.6\:\textrm{eV}\: \hspace{0.3cm} \mathrm{at} \hspace{0.3cm}\:  n_e(\mathrm{on\,Earth}) = 0.03 \: \mathrm{a.u.}
\end{equation}
provided that
\begin{align*}
\mathrm{a.u.} \equiv (4\pi/3)^{-1}\,a_\mathrm{B}^{-3}
\end{align*}
$Z_{eff}$ is tabulated as follows [26]:
\begin{align*}
\textrm{C}: 2.97 \: , \textrm{O}: 3.61\:, \textrm{Mg}: 6.39\: , \textrm{Si}: 7.66\,
\end{align*}
The enhanced C+C$\rightarrow$Mg and C+O$\rightarrow$Si pycnonuclear fusion reactions that take place in dense carbon-oxygen matters in a white dwarf cause enormous chemical potential change $\bar{\mu}$, that is
\begin{align*}
- \Delta G_r(\mathrm{C+C})=\phi^*(\mathrm{Mg})-2\phi^*(\mathrm{C})=\big\{Z_{\mathrm{eff}}(\mathrm{Mg})-2Z_{\mathrm{eff}}(\mathrm{C})\big\}\:\bar{\mu}
\end{align*}
\begin{equation}
=0.45\:\bar{\mu}
\end{equation}
\begin{align*}
- \Delta G_r(\mathrm{C+O})=\phi^*(\mathrm{Si})-\phi^*(\mathrm{C})-\phi^*(\mathrm{O}) 
\end{align*}
\begin{equation}
\,\,\,\,\,\,\,\,\,\,\,\,\,\,\,=\big\{Z_{\mathrm{eff}}(\mathrm{Si})-Z_{\mathrm{eff}}(\mathrm{C})-Z_{\mathrm{eff}}(\mathrm{O})\big\}\:\bar{\mu}=1.08\:\bar{\mu}\,
\end{equation}
The values of Eqs. (38) and (39) are applicable to the rough rate estimation of fusion in high-density plasma in a white dwarf through the correction of $\bar{\mu}$ in Eq. (37). Thus
\begin{equation}
\bar{\mu} (\mathrm{W.D})= \big\{\frac{n_e(\mathrm{W.D.})}{n_e(\mathrm{on \:Earth})}\big\}^{\frac{2}{3}} \bar{\mu}\: (\mathrm{on\: Earth}) = 10.7\times10^6\:\mathrm{eV}
\end{equation}
Here $n_e(\mathrm{W.D.})=2.5\times10^8\:\mathrm{a.u.}$ has been assumed [14]. The value of Eq. (40) is overestimated because $\bar{\mu}$ varies in proportion to $n_e^{\frac{1}{3}}$ rather than $n_e^{\frac{2}{3}}$ due to the relativistic effect of those extremely high-density electrons [14]. We prefer an intermediate $n_e$ dependence of $\bar{\mu}$, thus
\begin{align*}
\bar{\mu} (\mathrm{W.D})= \big\{\frac{n_e(\mathrm{W.D.})}{n_e(\mathrm{on \:Earth})}\big\}^{\frac{5}{9}} \bar{\mu}\: (\mathrm{on\: Earth}) = 7.5\times10^5\:\mathrm{eV} \tag{40'}
\end{align*}
As such,
\begin{equation}
-\Delta G_r(\mathrm{C+C}) \sim 3.5\times10^5\:\mathrm{eV}
\end{equation}
\begin{equation}
\mathrm{logK}(\mathrm{C+C}) \sim 18\:at\:T=10^8\:\mathrm{K}
\end{equation}
and
\begin{equation}
-\Delta G_r(\mathrm{C+O}) \sim 8.9\times10^5\:\mathrm{eV}
\end{equation}
\begin{equation}
\mathrm{logK}(\mathrm{C+O}) \sim 44 \:at\:T=10^8\:\mathrm{K}
\end{equation}
The first-principles calculation [42] is
\begin{equation}
\mathrm{logK}(\mathrm{C+C})=23.49\:at\:T=10^8\:\mathrm{K}
\end{equation}
and
\begin{equation}
\mathrm{logK}(\mathrm{C+O})=29.95 \: at \: T=10^8\:\mathrm{K}
\end{equation}

These values are not far from their counterparts in Eqs. (42) and (44). In conclusion, super enhanced chemonuclear fusion in liquid metals and enhanced pycnonuclear fusion in astrophysical condensed plasma are essentially in the same class. The only difference is the coherent H-H fusion within the correlating hydrogen ion pairs in the same sites of hydrogen adsorbent nanoclusters.

\hyphenchar\font=-1
\section*{1.6 Concluding Remarks}
\noindent
In the system of metallike hydride (e.g., NiD or PdD)--dense electron donor mixtures, adsorbed hydrogen atoms are transformed into metallic hydrogen around the octahedral sites of Ni or Pd lattices. Dense itinerant s-electrons supplied by the donor cause a drop in the melting point of the metallic hydrogen and induce their liquefaction. In the presence of thermodynamical liquid activity, hydrogen atoms become strongly correlated with the surrounding bulk of atoms, hence macroscopic scale collectivity. The coherent $\mathrm{D_2}$-$\mathrm{D_2}$ and $\mathrm{D_3}$-$\mathrm{D_3}$ fusion reactions at $T=460$ K -- enhanced with factors over $10^{20}$$\sim10^{30}$ and $10^{30}$$\sim10^{46}$ respectively -- take place, producing $\alpha$-particles of 23.8 MeV kinetic energy. These $\alpha$-particles facilitate the chemonuclear fission of non-fertile elements and the super enhanced disintegration of radioactive elements, among other things. We will discuss these extraordinary chemonuclear reactions in more detail in subsequent chapters.

\section*{References}
\begin{enumerate}
\small
\item Ichimaru S, Kitamura H (1999) Pycnonuclear reactions in dense astrophysical and fusion plasmas. Phys Plasmas 6: 2649--2671
\item Cameron AGW (1959) Pycnonuclear reactions and nova explosions. Astrophys J 130:916--940
\item Jancovici B (1977) Pair correlation function in a dense plasma and pycnonuclear reactions in stars. J Stat Phys 17:357--370
\item Ichimaru S, Kitamura H (1996) Thermodynamic enhancement of nuclear reactions in dense stellar matter. Publ Astron Soc Jpn 48:613--618
\item Ichimaru S (2000) Radiative proton-capture reactions of high-Z nuclei in the sun and in liquid--metallic hydrogen. Phys Lett A 266:167--172
\item Ikegami H (2001) Buffer energy nuclear fusion. Jpn J Appl Phys 40:6092--6098
\item Ikegami H, Pettersson R (2002) Evidence of enhanced nonthermal nuclear fusion. Bulletin of Institute of Chemistry, Uppsala University. pp 31--41
\item Ikegami H, Watanabe T, Petterson R, Fransson K (2007) Ultradense nuclear fusion in metallic lithium liquid. Revision of the Swedish Energy Agency Document ER2006:42, Tokyo, pp 3-1--3-30
\item Ibid. pp 4-1--4-21
\item Ibid. pp 1-1--1-22
\item Ibid. pp 2-1--2-12
\item Ikegami H, Hirose T, Sakai M, Yamazaki T, Sugiyama K (1968) E$0$, M$1$ and ($\mathrm{2_2}^{+}\rightarrow\mathrm{2_1}^{+}$) transitions in $^{188,190}\mathrm{Os}$, $^{192,194,196}\mathrm{Pt}$ and $^{198}\mathrm{Hg}$ Nuclei$^{+}$. In: Sanada J (ed) Proceedings of the international conference on nuclear structure, Tokyo, 1967. J Phys Soc Jpn 24 Suppl:167--171
\item Widom B (1963) Some topics in the theory of fluids. J Chem Phys 39:2808--2812
\item Ichimaru S (1993) Nuclear fusion in dense plasmas. Rev Mod Phys 65:255--299
\item Ikegami H, Pettersson R, Einarsson L (2004) Enormous entropy enhancement revealed in linked nuclear and atomic Li + D fusion in metallic Li liquid. Prog Theo Phys Suppl 154:251--260
\item Aylward G, Findlay T (1998) SI chemical data, 3rd edn. Wiley, Brisbane
\item Kondepudi D, Prigogine I (1998) Modern thermodynamics. Wiley, Chichester
\item Fukai Y (2005) The metal--hydrogen system: basic bulk properties, 2nd edn. Springer, Berlin Heidelberg
\item Wigner E, Seitz F (1933) On the constitution of metallic sodium. Phys Rev 43:804--810
\item Fowler WA, Caughlan GR, Zimmerman BA (1967) Thermonuclear reaction rates. Ann Rev Astron Astrophys 5:523--570
\item Krauss A, Becker HW, Trautvetter HP, Rolfs O, Brand K (1987) Low-energy fusion cross sections of D + D and D + $^{3}$He reactions. Nucl Phys A 465:150--172
\item Miedema AR (1973) The electronegativity parameter for transition metals: heat of formation and charge transfer in alloys. J Less-Common Metals 32:117--136
\item Rutherford E (1911) LXXIX. The scattering of $\alpha$ and $\beta$-particles by matter and the structure of the atom. Phil Mag Ser 6, 21:125, 669--688
\item Bohr N (1913) I. On the constitution of atoms and molecules, Phil Mag Ser 6, 26:151, 1--25
\item Weller RA (1992) Ion-solid interaction. In: Parker SP (ed) McGraw-Hill encyclopedia of science and technology, 7th edn, Vol 9:384. McGraw-Hill, New York
\item Puska MJ, Nieminen RM, Manninen M (1981) Atoms embedded in an electron gas: immersion energies. Phys Rev B 24:3037--3047
\item Andersen HH, Bay HL (1974) Nonlinear effects in heavy‐ion sputtering. J Appl Phys 45: 953--954
\item Ikegami H (1999) Coherent positronium molecule Ps$_\textit{n}$ and scanning clustering microscopy. Int J Quant Chem 71:83--99
\item Fukai Y, Sugimoto H (1985) Diffusion of hydrogen in metals. Adv in Phys 34:263--326
\item Nordlander P, Nørskov JK, Besenbacher F (1986) Trends in hydrogen heats of solution and vacancy trapping energies in transition metals. J Phys F 16:1161--1172
\item Sandrock G, Suda S, Schlapbach L (1992) Applications. In: Schlapbach L (ed) Hydrogen in intermetallic compounds II: surface and dynamic properties, applications. Springer, Berlin Heidelberg, pp 197--258
\item Boom R, de Boer FR, Miedema AR (1976) On the heat of mixing of liquid alloys--II. J Less-Common Metals 46:271--284
\item Kaufmann EN, Vianden R, Chelikowsky JR, Phillips, JC (1977) Extension of equilibrium formation criteria to metastable microalloys. Phys Rev Lett 39:1671--1674
\item Ikegami H (2012) The nature of the chemonuclear transitions. The Svedberg Laboratory, Uppsala University
\item A. Rossi, Communication to H. Ikegami, S. Kullander and R. Petter-\\sson 2012.
\item Essén H, Kullander S (2011) Experimental test of a mini-Rossi device at the Leonardo Corp, Bologna 29 March 2011. 
\item Sugimoto H, Fukai Y (1992) Solubility of hydrogen in metals under high hydrogen pressures: thermodynamical calculations. Acta Metallurgica et Materialia 40:2327--2336
\item Völkl J, Alefeld G (1978) Diffusion of hydrogen in metals. In: Alefeld G, Völkl J (eds) Hydrogen in metals I: basic properties. Springer, Berlin Heidelberg, pp 321--348
\item Wipf H (1997) Diffusion of hydrogen in metals. In: Wipf H (ed) Hydrogen in metals III: properties and applications. Springer, Berlin Heidelberg, pp 51--91
\item Conrad H, Ertl G, Latta EE (1974) Adsorption of hydrogen on palladium single crystal surfaces. Surf Sci 41:435--446
\item de Ninno A, Frattolillo A, Rizzo A, Giudice DE, Preparata G (2002) Experimental evidence of $^{4}$He production in a cold fusion experiment. RT/2002/41/FUS, Servizio Edizioni Scientifiche -- ENEA Centro Ricerche Frascati, Rome
\item Ogata S, Iyetomi H, Ichimaru S (1991) Nuclear reaction rates in dense carbon-oxygen mixtures. Astrophys J 372:259--266
\item Ikegami H (1990) Cyclotron maser cooling of electron and ion beams. Phys Rev Lett 64: 1737--1740

\end{enumerate}
\end{sloppypar}
\end{document}